\documentstyle[sprocl]{article}

\def\Journal#1#2#3#4{{#1} {\bf #2}, #3 (#4)}

\def\NCA{\em Nuovo Cimento}

\def\NPB{{\em Nucl. Phys.} B}
\def\PLB{{\em Phys. Lett.}  B}

\def\PRD{{\em Phys. Rev.} D}
\def\ZPC{{\em Z. Phys.} C}

\def\als{\alpha_{\rm s}}
\def\lQ{\Lambda_{\rm QCD}}
\def\siml{{\ \lower-1.2pt\vbox{\hbox{\rlap{$<$}\lower6pt\vbox{\hbox{$\sim$}}}}\ }} 

\def\be{\begin{equation}}
\def\ee{\end{equation}}
\def\bea{\begin{eqnarray}}
\def\eea{\end{eqnarray}}

\begin{document}

\title{The Role of the QCD Vacuum in the Heavy-Quark Bound State Dynamics\footnote{
Talk given by N.B. at the Fifth Workshop on Quantum 
Chromodynamics, Villefranche-sur-Mer, France, 3-7 January 2000.} }

\author{NORA BRAMBILLA and  ANTONIO VAIRO}

\address{Institut f\"ur Theorestische Physik, Universit\"at Heidelberg\\
Philosophenweg 16, D-69120 Heidelberg, Germany \\
and\\
Theory Division CERN, 1211 Geneva 23, Switzerland
\\E-mail: nora.brambilla@cern.ch, antonio.vairo@cern.ch} 


\maketitle\abstracts{The effective field theory approach allows a rigorous
disentangling of high and low energy  effects in the heavy quarkonium dynamics. 
Focusing in particular on the spectrum, we describe the nature of the non-perturbative 
effects and discuss our present knowledge of them.}

\section{Introduction}
It is well known that the complex structure of the QCD vacuum can, in some 
regimes, be parametrized by local vacuum condensates, i.e. the expectation values of operators 
where all the perturbative vacuum fluctuations are taken out at the best of
our ability \cite{sumrules}.
The first attempt to calculate the effect of non-perturbative vacuum condensates  
on the energy levels of heavy quarkonia was performed in \cite{vole}. 
The leading contribution (in $\als$ and $\lQ$, the scale of non-perturbative 
physics) of the  vacuum condensates to Coulombic $n,l$ heavy quarkonium states reads  \cite{vole}:
\begin{equation}
\delta E^{\rm V-L}_{nl} = m {\epsilon_n  n^6  \pi^2  G_2 \over (m C_F \als)^4},   
\label{contvole}
\end{equation}
where $\epsilon_n$ is a (known) number of order 1 and $G_2\equiv \langle (\als
/ \pi) F^a_{\mu\nu}(0) F^{a\,\mu\nu}(0) \rangle $ is the gluon condensate.
The above expression for the corrections to Coulombic energy levels displays
two relevant characteristics: 1) it is a correction of non-potential type, like the  
Lamb shift in QED; 2) it grows like $\sim  n^6$, thus being out of control for levels 
beyond the ground state. It was soon realized \cite{nonlocal,mar} that the strong growth in $n$ 
could be corrected to some extent by considering non-local gluon condensates  
\begin{equation}
G_2(x)  \sim \left\langle  {\als\over \pi} F^a_{\mu\nu}(x)  \phi(x,0)^{\rm
    adj}_{ab} F^{b\,\mu\nu}(0) \right \rangle, 
\end{equation}
where $\phi(x,0)^{\rm adj}$ is the Schwinger line (in the adjoint representation) connecting $x$ with $0$.
These were understood as due to the presence of a fluctuating gluonic background with a 
characteristic time length $T_g \sim \lQ^{-1}$. It is apparent that the local condensates stem 
from an expansion of the non-local ones in cases in which the correlation
length is large with respect to the other physical scales of the system.  

Heavy quarkonium is a non-relativistic bound system. Besides $\lQ$, it is characterized 
by at least three hierarchically ordered scales, $m v^2 \ll m v \ll m$, where
$m$ is the mass of the heavy quark and $v$ its velocity.  
Therefore, it is only under the condition $\lQ \ll m v^2 $ that the use 
of local condensates and, hence, the Voloshin--Leutwyler formula (\ref{contvole}) can be justified. 
In the other regimes where heavy quarkonium states can sit, the non-perturbative dynamics 
will be encoded into more extended objects: non-local condensates and Wilson loop operators.
This is the case for higher quarkonium levels ($n>1$).    

In the following we will consider non-perturbative effects in heavy quarkonium
systems in different kinematic regimes, in an effective field theory context. 
This approach has not only the considerable practical advantage of disentangling the different 
dynamical scales of the system, but also the conceptually relevant feature of disentangling 
perturbative effects from non-perturbative ones. This allows us to fully exploit the 
predictive power of QCD.

\section{Local and non-local condensates: $\lQ \ll mv$}
For the lower lying quarkonium states, it can be expected that the inverse of the typical size 
of the system is larger than $\lQ$. If this condition is fulfilled, both scales $m$ and $mv$ can be 
integrated out perturbatively from QCD, leading to an effective field theory where 
only the degrees of freedom of order $\lQ$ or $mv^2$ remain dynamical. 
This effective field theory is known as potential NRQCD, pNRQCD \cite{pnrqcd0,pnrqcd1}.

The infrared sensitivity of the quark--antiquark static potential at three loops \cite{pot} signals that 
it may become sensitive to non-perturbative effects if the next relevant scale 
after $m v$ is $\lQ$. Indeed, in the situation $mv \gg \lQ \gg mv^2$, the leading non-perturbative 
contribution (in $\als$ and in the multipole expansion) to the static potential reads \cite{pnrqcd1,balitsky}
\begin{equation}
V_0(r)^{\rm non-pert}  
= -i{g^2 \over N_c} T_F {r^2\over 3} \int_0^\infty \!\! dt \, e^{-itC_A \als /(2 r)} 
\langle {{\bf E}^a}(t) \phi(t,0)^{\rm adj}_{ab} {\bf E}^{b}(0) \rangle(\mu), 
\label{potnon}
\end{equation}
where $r$ is the quark--antiquark distance.
This term explicitly cancels, up to the considered order, the dependence of the perturbative 
static potential on the infrared scale $\mu$. It is interesting to note that the leading 
contribution in the $\lQ/mv^2$ expansion of $V^{\rm non-pert}$ (obtained by
putting the exponential equal to 1) cancels the order $\lQ^3 r^2$ renormalon
that affects the static potential (the leading-order renormalon, of order $\lQ$, cancels against the 
pole mass). Therefore, also in renormalon language, the above operator is the relevant 
non-perturbative contribution to the static potential in the considered kinematic situation.

If $\lQ \siml mv^2$ the static potential is purely perturbative and its explicit dependence 
on the infrared scale $\mu$ is reabsorbed in a physical observable by non-potential 
contributions. In the specific case of the quarkonium  energy levels up to order 
$\als^5 \ln \mu$, these contributions are \cite{lev}
\begin{eqnarray}
& & \!\!\!\!\!\!\!\! \delta E_{n,l,j}  = -i{g^2 \over 3 N_c}T_F  \int_0^\infty \!\! dt \,
\langle n,l |  {\bf r} e^{it( E_n - H_o)} {\bf r }  | n,l \rangle 
\langle {\bf E}^a (t) \phi(t,0)^{\rm adj}_{ab} {\bf E}^b (0) \rangle (\mu),
\label{Engen} \\
& & \!\!\!\!\!\!\!\!  E_n  \equiv -  m  {C_F^2 \alpha_{\rm s}^2  \over 4  n^2}, 
\qquad H_o \equiv {{\bf p}^2\over m} + {1\over 2 N_c}{\als \over r}, 
\nonumber
\end{eqnarray}
where $ | n,l \rangle $ are the Coulomb wave functions. It is worth while to notice that, 
if $\lQ \ll mv^2$, the scale $m v^2$ can be integrated out perturbatively from
the above formula and the non-perturbative contributions reduce to the
Voloshin--Leutwyler formula (\ref{contvole}), as can easily be seen by recognizing that:
$$
m {\epsilon_n  n^6  \over (m C_F \als)^4} 
= {1 \over 3 N_c} T_F \left\langle n,l \left|  {\bf r} {1 \over E_n - H_o }
    {\bf r }  \right| n,l \right\rangle. 
$$

The above outline leads to the following conclusion. For quarkonium of a typical size 
smaller than $1/\lQ$, the most relevant operator of the non-perturbative dynamics is the bilocal gluon condensate 
$\langle {\bf E}^a (t) \phi(t,0)^{\rm adj}_{ab} {\bf E}^b (0) \rangle $, which belongs to the class 
of non-local gluon condensates considered in the introduction. In the following section 
we will discuss our present knowledge of it.

\subsection{The non-local condensate $\langle F^a_{\mu\nu} (x) \phi(x,0)^{\rm adj}_{ab} F^b_{\lambda\rho} (0) \rangle$}
\label{correlator}
The correlator $\langle F^a_{\mu\nu} (x) \phi(x,0)^{\rm adj}_{ab} F^b_{\lambda\rho} (0) \rangle$ 
is perturbatively known at the next-to-leading order in $\als$ \cite{eeper}. 
However,~ here we are ~interested in ~its non-perturbati\-ve ~behaviour. 
Different ~parametrizations have been ~proposed 
\cite{sv,kra,hyb,lat1,lat2}. Because of its Lorentz structure, the correlator is in general
described by two form factors. A convenient choice of these
consists in the chromoelectric and chromomagnetic correlators:
$$
\langle {\bf E}^a (x) \phi(x,0)^{\rm adj}_{ab} {\bf E}^b (0) \rangle\, ,
\qquad\qquad
\langle {\bf B}^a (x) \phi(x,0)^{\rm adj}_{ab} {\bf B}^b (0) \rangle.
$$
The strength of the correlators is of the order of the 
gluon condensate. In the long range ($x^2 \to \infty$) they fall
off exponentially (in the Euclidean space) with some typical correlation lengths. 
In the following we will concentrate on these correlation lengths.

The lattice calculation \cite{lat1}, using cooling techniques, 
obtains the same correlation length ($T_g$) for both form factors, and this is 
\begin{eqnarray}
& &T_g = 0.34 \pm 0.02 \pm 0.03 \hbox{~fm ~(4 flavours,~}  a m =  0.01 \hbox{)},
\label{latt1}\\
& &T_g = 0.22 \pm 0.01 \pm 0.02 \hbox{~fm ~(quenched)}.
\label{latt2}
\end{eqnarray}
The less accurate, but traditional (quenched) lattice calculation done in
\cite{lat2} obtains two different correlation lengths for the 
chromoelectric ($T_g^E$) and the chromomagnetic correlators ($T_g^B$):
\begin{eqnarray}
T_g^E \neq T_g^B \simeq 0.1\hbox{--}\,0.2 \hbox{~fm ~(quenched)}.
\label{latt3}
\end{eqnarray}
Finally, a recent sum-rule estimation \cite{hyb} obtains $T_g^E < T_g^B$. 
The sum rule turns out not to be stable for the chromoelectric correlator, 
while for the chromomagnetic correlation length it gives  
\begin{eqnarray}
& &{T_g}^B  = 0.13^{+0.05}_{-0.02} \hbox{~fm ~(3 flavours)},
\label{sum1}\\
& &{T_g}^B  = 0.11^{+0.04}_{-0.02} \hbox{~fm ~(quenched)}.
\label{sum2}
\end{eqnarray}

The correlation lengths $T_g^E$ and $T_g^B$ have a precise physical
interpretation. Their inverses correspond to the masses of the 
lowest-lying vector and pseudovector static quark--gluon hybrids, respectively. 
This can be explicitly seen in the short-range limit, ${\bf x} \to 0$, 
where the hybrids (in this case also called gluelumps) operators can be 
explicitly constructed \cite{pnrqcd1,brahyb}. The suitable effective field 
theory is pNRQCD in the static limit \cite{pnrqcd1}. Gluelump operators
are of the type ${\rm Tr}\{{\rm O}H\}$, where ${\rm O} = O^aT^a$ corresponds 
to a quark--antiquark state in the adjoint representation (the octet) and 
$H = H^a T^a$ is a gluonic operator. By matching the QCD static hybrid 
operators into pNRQCD, we get the static energies (also called potentials) 
of the gluelumps. At leading order in the multipole expansion, they read
\begin{eqnarray}
& & V_H(r) = V_o(r) + {1 \over T_g^H},
\label{potglue}\\
& & \langle H^a(t) \phi(t,0)^{\rm adj}_{ab}H^b(0)\rangle^{\rm non-pert.} 
\simeq h \, e^{- i t/T_g^H} + \dots .
\nonumber
\end{eqnarray}
Since hybrids are classified in QCD according to the representations of
$D_{\infty,h}$, while in pNRQCD, where we have integrated out the length $r$, 
their classification is done 
according to the representations of $O(3)\times C$, the static hybrid
short-range spectrum is expected to be more degenerate than the long-range one  
\cite{michael,brahyb}. The lattice measure of the hybrid potentials 
done in \cite{Morningstar} confirms this feature. In \cite{pnrqcd1} it has
been shown that the quantum numbers attribution of pNRQCD to the short-range
operators, and the expected $O(3) \times C$ symmetry of the effective field 
theory match the lattice measurements. By using only ${\bf E}$ and
${\bf B}$ fields and keeping only the lowest-dimensional representation 
we may identify the operator $H$ for the short-range hybrids called $\Sigma_g^{+\,'}$ 
(and $\Pi_g$) with ${\bf r}\cdot{\bf E}$ (and ${\bf r}\times{\bf E}$) 
and the operator $H$ for the short-range hybrids called $\Sigma_u^{-}$ 
(and $\Pi_u$) with ${\bf r}\cdot{\bf B}$ (and ${\bf r}\times{\bf B}$). 
Hence, the corresponding static energies for small $r$ are
$$
V_{\Sigma_g^{+\,'},\Pi_g}(r) = V_o(r) + {1 \over T_g^E}, \qquad\qquad
V_{\Sigma_u^{-},\Pi_u}(r) = V_o(r) + {1 \over T_g^B}.
$$
The lattice measure of \cite{Morningstar} shows that, in the short range, 
$\!V_{\Sigma_g^{+\,'},\Pi_g}(r)\! >\! V_{\Sigma_u^{-},\Pi_u}(r)$. 
This supports the sum-rule prediction \cite{hyb} that the pseudovector 
hybrid lies lower than the vector one, i.e. $T_g^E < T_g^B$.

\section{Wilson loop operators: $\lQ \sim mv$}
For higher quarkonium levels, $\lQ$ is expected to be comparable with $mv$.
We cannot match into pNRQCD perturbatively, since the scale associated 
to the quarkonium size $r$ is already non-perturbative. The 
relevant non-perturbative dynamics is therefore contained in more extended objects than 
(local or non-local) gluon condensates: Wilson loops and field insertions on these.

In particular, disregarding effects due to scales lower than $\lQ$, the
static potential is given by \cite{wilson}
\begin{equation}
V_0(r) = \lim_{T\to\infty}{i\over T} \ln \langle W_\Box \rangle, 
\label{v0}
\end{equation}
where $W_\Box$ is the static Wilson loop of size $r \times T$ and 
$\langle ~~\rangle$ means an average over the gauge fields. 
Lattice studies tell us that at distances $r \simeq 1/\lQ$ the potential 
is no longer  Coulombic but rises linearly ($V_0(r) \simeq \sigma r$).  
Higher-order corrections in the $1/m$ expansion have been calculated over the years \cite{horder} 
and are given by field strength insertions on the Wilson loop. For instance the next-to-leading
potential in the $1/m$ expansion is \cite{m1}
\begin{eqnarray}
{V_1\over m} &=& {1\over m} \lim_{T\to\infty}\Bigg(
- {g^2\over 4 T}\int_{-T/2}^{T/2} \!\! dt \int_{-T/2}^{T/2} \!\!dt^\prime
\vert t -t^\prime \vert \bigg[ \langle\!\langle {\bf E}(t) \cdot {\bf E}(t^\prime)\rangle\!\rangle_\Box 
\\
& & \qquad\qquad\qquad\qquad\qquad\qquad
- \langle\!\langle {\bf E}(t)\rangle\!\rangle_\Box \cdot
\langle\!\langle {\bf E}(t^\prime)\rangle\!\rangle_\Box  \bigg] \Bigg),
\label{v1E}
\end{eqnarray}
where $\langle\!\langle ~~\rangle\!\rangle_\Box$ means a normalized gauge average 
in the presence of the static Wilson loop. 

Wilson-loop operators of the above type may be interpreted as a superposition of states, 
describing gluonic excitations between static sources \cite{m1}. 
They have been so far evaluated only inside QCD vacuum models or 
by lattice simulations (for some reviews, see \cite{rev} and also \cite{mod}).

\subsection{The stochastic expansion}
We can now ask if some relation can be established between the Wilson loop (and field
strength insertions on it) and the non-local gluon condensates discussed
above. In fact we can express the Wilson loop as a formal expansion in terms  
of gluonic correlation functions by means of the so-called stochastic
expansion \cite{kampen,sv} 
\begin{eqnarray}
\ln \langle W_\Box \rangle &=& \sum_{n=0}^\infty {(ig)^n \over n!}
\int_{S(\Box)} dS_{\mu_1\nu_1}(u_1) \cdots  dS_{\mu_n\nu_n}(u_n) \langle \phi(0,u_1)
\nonumber\\
&~& \times F^{\mu_1\nu_1}(u_1)\phi(u_1,0) \cdots \phi(0,u_n) F^{\mu_n\nu_n}(u_n)\phi(u_n,0) 
\rangle_{\rm cum}, 
\label{cumex}
\end{eqnarray}
where $S(\Box)$ denotes a surface whose contour the rectangular Wilson loop.
The cumulants $\langle ~~ \rangle_{\rm cum}$ are defined as
\begin{eqnarray*}
& & \langle \phi(0,u_1) F(u_1) \phi(u_1,0) \rangle_{\rm cum} 
\quad = \langle \phi(0,u_1) F(u_1) \phi(u_1,0) \rangle = 0 , \\
& & \langle \phi(0,u_1) F(u_1) \phi(u_1,u_2) F(u_2) \phi(u_2,0)\rangle_{\rm cum} = \\
& &\qquad\>\> \langle \phi(0,u_1) F(u_1) \phi(u_1,u_2) F(u_2) \phi(u_2,0)\rangle  \\
& &\qquad  - \langle \phi(0,u_1) F(u_1) \phi(u_1,0)\rangle\,  
\langle \phi(0,u_2) F(u_2)\phi(u_2,0)\rangle \dots  = \\ 
& & \langle F(u_1) \phi(u_1,u_2)^{\rm adj} F(u_2) \rangle , \\
&\,& \qquad\qquad\qquad\qquad\qquad \cdots \quad \quad \cdots \nonumber  
\end{eqnarray*}
It is important to realize that the expansion of Eq. (\ref{cumex}) is
substantially different with respect to the expansions in $1/m$ and $r$, 
which led to the construction of the low-energy effective field theories discussed above.  
Those expansions were justified by the dynamics of the system under study, and
each term of it is, indeed, suppressed by powers of the highest dynamical scale
left divided by the scale (which is larger) that has been integrated out. 
Instead, from a power counting point of view, each term of the expansion
(\ref{cumex}) is of the same size (for instance each term of the 
series is in general expected to contribute to the string tension $\sigma$). 
As soon as we truncate the series, e.g. up to the bilocal cumulant (the first non-vanishing one), 
we introduce an uncontrolled approximation and define a model. This model is
known as the model of the stochastic vacuum \cite{sv}.
Indeed, this model turns out to be quite successful in the study of processes 
that involve quark systems that may be described by almost static Wilson loops 
(for some reviews, see \cite{svmrev}). Let us only mention that the model predicts, 
in agreement with the lattice data, a long-range 
linear static potential with slope $\sigma \sim T_g^{E\,2} G_2$. 
It would be highly desirable to have a field theoretical justification 
of the expansion (\ref{cumex}); however, such a justification is missing up to
now (for a recent investigation on higher cumulants, see \cite{Kor}).

\section{Conclusions}
In the previous sections we have shown how the non-perturbative QCD vacuum 
enters the dynamics of heavy quarkonium. As much as the non-perturbative
scale $\lQ$ is bigger than the dynamical scales of the non-relativistic system 
($m$, $mv$ and $m v^2$), as extended the relevant non-perturbative operators are. 
In the situation $\lQ \ll m v^2$ these operators reduce to local condensates, i.e. some
numbers. This is the most favourable situation for a theoretical
investigation. The bottomonium and charmonium ground states have been investigated 
in this framework (for a recent review, see \cite{revqq}). In the limiting case 
of $t\bar{t}$ threshold production the non-perturbative corrections can be
neglected and the investigation is completely accessible to perturbative QCD \cite{hoang}.
In the situation $m v^2 \siml \lQ \ll m v$ the non-perturbative physics is
encoded into non-local condensates. The dominant one is the bilocal gluon
condensate. As we have discussed above it essentially
depends on its strength, i.e. the gluon condensate, and on some 
correlation lengths, which can be related to the masses of the lowest-lying
(static) quark--gluon hybrid resonances.  Although this situation seems quite
interesting, it has been poorly investigated in heavy quarkonium phenomenology, and
mainly in the framework of models \cite{balitsky,mar}. The main reason is, in our
opinion, that, while the quarkonium ground state seems to be accessible by a
purely perturbative treatment (plus local condensates) and the potential 
describing higher excited quarkonium states seems entirely dominated by 
non-perturbative effects, it is, a priori, not clear to which states the 
situation $m v^2 \siml \lQ \ll m v$ is applicable. We will come back 
to this point at the end of this section. 
Finally, in the situation $\lQ \sim mv$ the relevant non-perturbative 
objects are Wilson-loop operators. A lattice study of them and the subsequent 
quarkonium spectroscopy have been done in \cite{bal97}. 

The outlined study is rigorous and allows a systematic disentanglement
of the high from the low energy scales of the heavy quarkonium system under
study. In this specific sense also perturbative and non-perturbative effects 
turn out to be disentangled. 

We conclude by mentioning a somehow delicate point, connected with the scales 
of heavy quarkonium systems: the difficulty to state {\it a priori} in which 
kinematic situation a particular quarkonium state is. This is mainly due to the fact that 
the scales $m v$ and $m v^2$ are not so well defined, nor so widely separated that different 
kinematic situations cannot overlap in a preliminary analysis (for a related  discussion on 
the energy scales also in quarkonium physics, see \cite{shu}). Therefore, there are situations 
in which scales can only be fixed {\it a posteriori}, i.e. by assuming a particular situation and 
by checking that the final result is consistent with it and with the experimental data. 

\vspace{0.8cm}

\noindent {\bf Acknowledgements}\vspace{3mm} \\ \noindent 
N.B. would like to thank for the invitation Jan Rafelski, chairman and organizer of the Stochastic 
Vacuum Session and the organizers of the conference, Herbert Fried, Bernd M\"uller and Yves Gabellini.
N.B. is grateful to the theory group of the University of Milan for supporting
her participation to the conference. N.B. and A.V. gratefully acknowledge the 
Alexander von Humboldt Foundation.

\end{document}